%% file: ms.tex
\documentstyle[psfig,aaspp4]{article}

\begin{document}

\title{M DWARFS FROM {\it HUBBLE SPACE TELESCOPE} STAR COUNTS. IV.
\footnote{Based on observations with the NASA/ESA Hubble Space Telescope,
obtained at the Space Telescope Science Institute, which is operated by 
the Association of Universities for Research in Astronomy, Inc. under 
NASA contract No. NAS5-26555.}}

\author{Zheng Zheng\altaffilmark{2}, Chris Flynn\altaffilmark{3},
        Andrew Gould\altaffilmark{2,4}, John N. Bahcall\altaffilmark{5}, 
        Samir Salim\altaffilmark{2}}

\altaffiltext{2}{Department of Astronomy, Ohio State University, Columbus, 
                 OH 43210; zhengz@astronomy.ohio-state.edu, 
                 gould@astronomy.ohio-state.edu, 
                 samir@astronomy.ohio-state.edu}
\altaffiltext{3}{Tuorla Observatory, Turku University,
                 V\"{a}is\"{a}l\"{a}ntie 20, FIN-21500, Piikki\"{o},
                 Finland; cflynn@astro.utu.fi}
\altaffiltext{4}{Laboratoire de Physique Corpusculaire et Cosmologie,
                 Coll\`ege de France, 11 pl.\ Marcelin Berthelot, F-75231, 
                 Paris, France}
\altaffiltext{5}{Institute for Advanced Study, Princeton, NJ 08540; 
                 jnb@ias.edu}

\begin{abstract}
We study a sample of about 1400 disk M dwarfs that are found in 
148 fields observed with the Wide Field Camera 2 (WFC2) on the {\it 
Hubble Space Telescope} and 162 fields observed with pre-repair 
Planetary Camera 1 (PC1), of which 95 of the WFC2 fields are newly 
analyzed. The method of maximum likelihood is applied to derive the 
luminosity function and the Galactic disk parameters. At first, we use 
a local color-magnitude relation and a locally determined mass-luminosity 
relation in our analysis. The results are consistent with those of previous
work but with considerably reduced statistical errors. These small 
statistical errors motivate us to investigate the systematic uncertainties. 
Considering the metallicity gradient above the Galactic plane, we introduce 
a modified color-magnitude relation that is a function of Galactic height. 
The resultant M dwarf luminosity function has a shape similar to that 
derived using the local color-magnitude relation but with a higher peak 
value.  The peak occurs at $M_V \sim 12$ and the luminosity function drops 
sharply toward $M_V \sim 14$. We then apply a height-dependent
mass-luminosity function interpolated from theoretical models with
different metallicities to calculate the mass function. Unlike the mass 
function obtained using local relations, which has a power-law index 
$\alpha = 0.47$, the one derived from the height-dependent relations
tends to be flat ($\alpha = -0.10$). The resultant local surface density 
of disk M dwarfs ($12.2\pm1.6 M_\odot {\rm pc}^{-2}$) is somewhat smaller 
than the one obtained using local relations ($14.3\pm1.3 M_\odot 
{\rm pc}^{-2}$). Our measurement favors a short disk scale length, 
$H = 2.75 \pm 0.16 {\rm (statistical)} \pm 0.25 {\rm (systematic)}$ kpc. 
\end{abstract}


\section{INTRODUCTION}

M dwarfs dominate the luminous  matter in the disk of the Galaxy. Thus 
it is important to study M dwarfs in order to constrain the disk mass 
and to understand the spatial distribution of stars in the disk. The 
mass function of M dwarfs may also give us some hints about the number 
of brown dwarfs whose masses are below the hydrogen-burning limit. 
Moreover, M dwarfs contribute to the observed microlensing events.

Star counts provide a straightforward way to explore the above
questions. Work on star counts has a long history and developments 
of new techniques in detection and data reduction have made it a 
powerful tool to study the structure of the Galaxy (see Bahcall 1986 
for a review). During the last twenty years, a variety of efforts have 
been made to count M dwarfs with ground-based observations (e.g. Hawkins 
\& Bessel 1988; Tinney, Reid, \& Mould 1993; Kirkpatrick et al. 1994). 

Ground-based photometric studies and parallax studies generally deal 
with relatively nearby stars. Hence, they are not sensitive to the 
overall distribution of the stars in the Galactic disk and are subject 
to Malmquist bias. Observations made by the {\it Hubble Space Telescope} 
(HST) permit the resolution of much more distant stars. Hence, the overall 
distribution of stars in the disk can be measured more accurately. At the 
same time, Malmquist bias can be greatly reduced since the survey reaches 
the ``top" of the disk. The present study is the culmination of 
almost a decade of work on HST star counts, beginning with counts using 
the pre-repair Planetary Camera (PC1) (Gould, Bahcall, \& Maoz 1993) and 
the first much deeper counts with the repaired Wide Field Camera (WFC2) 
(Bahcall et al.\ 1994). Gould, Bahcall, \& Flynn (1996, hereafter Paper I) 
studied a sample of 257 Galactic disk M dwarfs which include 192 stars 
in 22 fields observed with WFC2 with mean limiting magnitude $I=23.7$ and 
65 stars in 162 fields observed with PC1 with mean limiting magnitude 
$V=21.3$. In this paper, $V$ and $I$ denote magnitudes in Johnson-Cousins 
systems. They derived a disk luminosity function (LF) peaking at 
$M_V \sim 12$ and dropping off sharply between $M_V=12$ and $M_V=14$. 
The total column density of M dwarfs at the Galactocentric radius 
$R_0=8.0$ kpc was determined to be $12.4\pm1.9 M_\odot {\rm pc^{-2}}$. 
The scale length for the M-star disk was found to be $3.0\pm0.4$ kpc. 
In a follow-up paper, Gould, Bahcall, \& Flynn (1997, hereafter Paper II) 
incorporated 80 additional M dwarfs found in 31 new fields observed with 
WFC2 into their data set. The overall results were consistent with Paper 
I but with somewhat smaller error bars. 

In this paper, we analyze the disk M dwarfs found in an additional 95 WFC2 
fields. After combining these 95 fields with the fields studied in Papers 
I and II, our sample of disk M dwarfs now includes about 1400 stars, almost 
three times larger than the sample studied in Paper II. The large sample 
helps to reduce the statistical uncertainties, especially at the faint end 
of the LF. Although the underlying method is the same as that used in 
Papers I and II, we make several modifications in this paper. First, we 
find that the errors of $V$ and $I$ magnitudes are slightly underestimated 
in previous work due to a bug in the computer code. We correct this bug.
For most of the stars, the correction is negligibly small. However, this 
correction has a relatively larger effect for the last $M_V$ bin. Second, 
in this paper, we adopt a slightly revised photometric transformation from 
WFC2 instrumental to standard Johnson-Cousins magnitudes, which is derived 
through an empirical calibration and is described in the Appendix. Third, 
we take the metallicity effect on the color-magnitude relation (CMR) into 
account. Based on the color-magnitude diagram by Reid (1991) and Monet et 
al. (1992), we add a term varying with the Galactic height $z$ which is 
designed to model the metallicity effect. We then  interpolate the 
mass-$M_V$ relations based on different metallicities 
(Baraffe et al.\ 1998) to derive the mass function, rather than simply 
using the local (solar metallicity) relation of Henry \& McCarthy (1993) 
as was done previously.
 
The main results of this paper are: First, if we use the original 
solar-neighborhood CMR (Reid 1991), the overall results are consistent 
with those in Paper I and II but with considerably smaller statistical 
errors. The M dwarf mass function has a power-law index $\alpha=0.47$
in the range of $\sim 0.08 M_\odot$ to $\sim 0.5 M_\odot$. Second, if we 
adopt the modified $z$-dependent CMR, the best fit scale heights and 
scale lengths in our models are about 20-30\% smaller. The local M dwarf 
mass density and surface density are about 15\% smaller than those using 
the solar-neighborhood CMR, and the mass function is roughly flat, 
$\alpha \sim -0.10$.

\section{OBSERVATIONS AND DATA ANALYSIS}

\subsection{Observations}

In this paper, we include three sets of data (data in Paper I, new data 
in Paper II, and new fields in this paper). Altogether we include 148
fields imaged with WFC2 and 162 fields imaged with PC1. The WFC2 fields
were chosen to satisfy the following criteria: first, the Galactic 
latitude $|b| > 17^\circ$; second, there should be at least 2 exposures 
with the F814W filter and at least one with F606W filter for each field; 
third, in these fields, there should be no Local Group galaxies or 
globular clusters in our Galaxy or other galaxies. The 162 PC1 fields are 
taken from 166 QSO snapshot survey fields (Bahcall et al.\ 1992; Maoz et 
al.\ 1993) for which the QSOs were selected by radio, X-ray, and 
color-excess techniques. Four PC1 fields were excluded because ground-based
observations could not be obtained.

We set two magnitude limits for each field observed with WFC2. The 
faint magnitude limit $I_{\rm max}$ denotes our detection threshold. 
This limit ensures that the discrimination between stars and galaxies
is clear. The bright magnitude limit $I_{\rm min}$ represents the 
saturation threshold. Details on these two limits can be found in 
Paper I.

The photometry for the stars in the 162 fields imaged with PC1 is based 
on ground observations (see Paper I). These stars with non-HST photometry 
occupy an important part of parameter space. Generally speaking, the 
stars found in these fields would have been saturated in WFC2 data because 
they have relatively small Galactic heights and high luminosities. Hence, 
these stars (denoted by open circles) provide much of the data occupying 
the lower-left part of the $z-M_V$ plane shown in Figure 1.

\begin{figure}[h]
\centerline{\psfig{figure=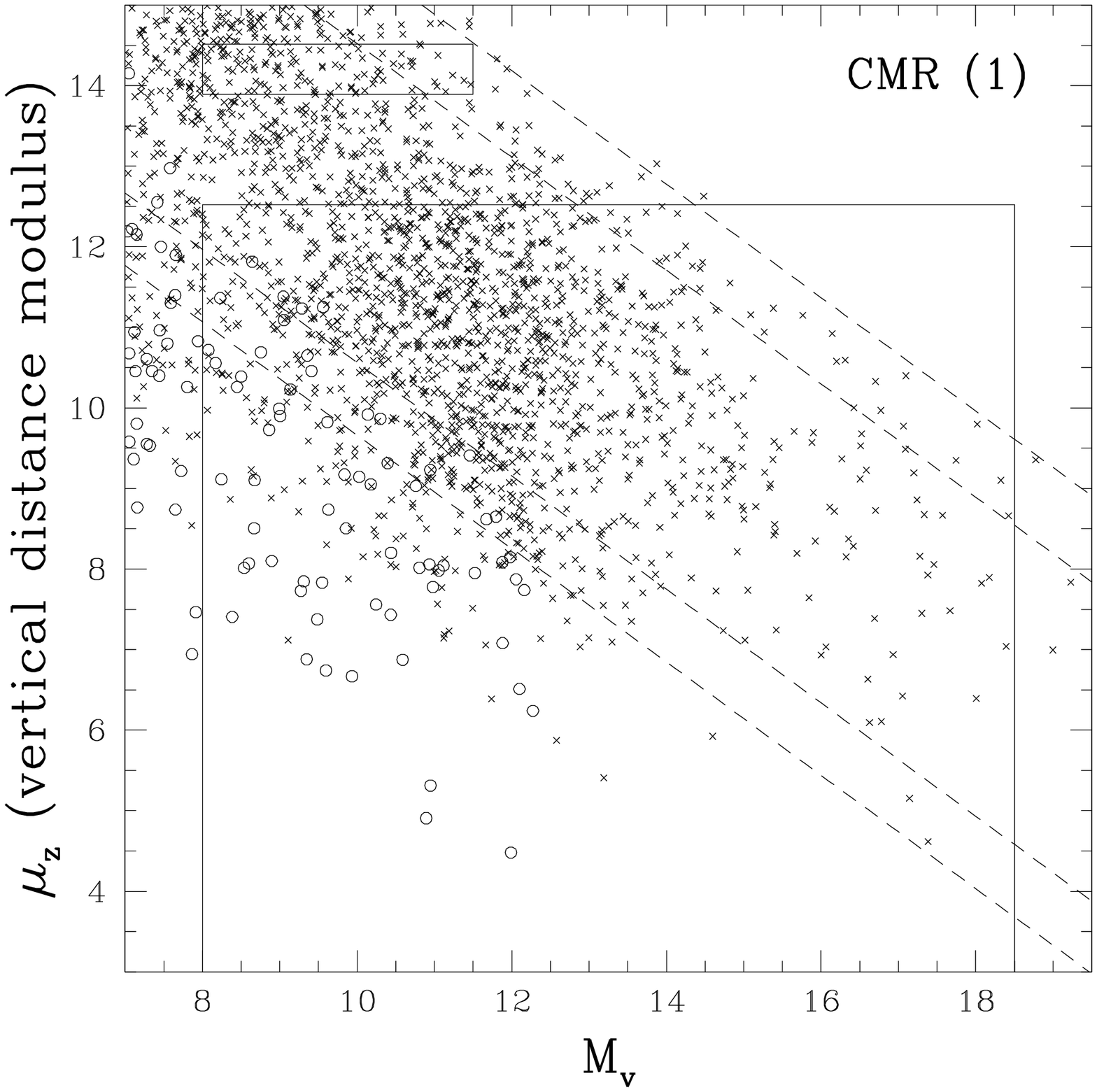,height=8cm,width=8cm}
            \psfig{figure=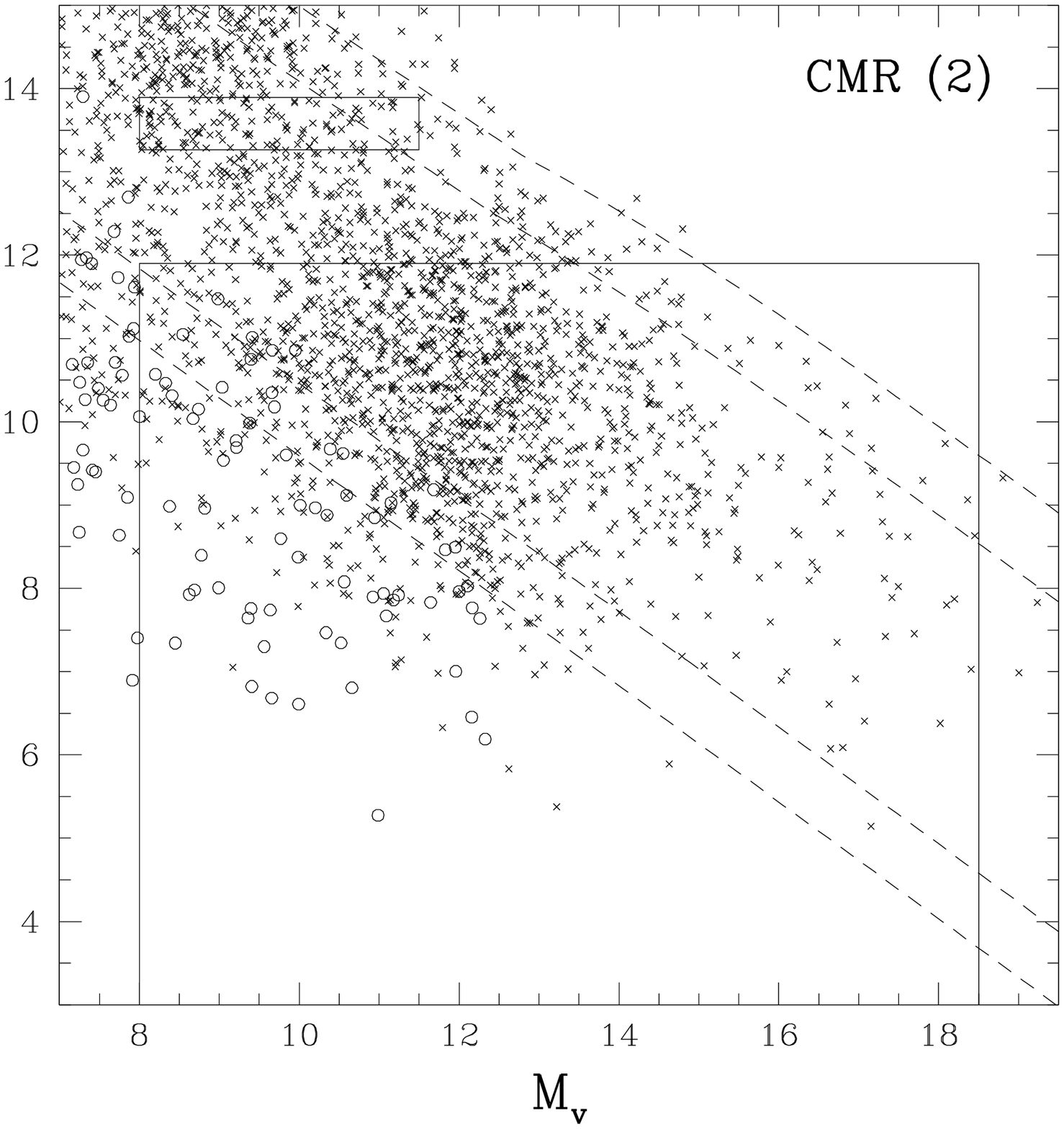,height=8cm,width=8cm}}
\caption[]{Stars in all the fields shown in the $\mu_{\rm z}-M_V$ plane,
where $\mu_{\rm z}=5\lg z -5 = V_0-M_V+5\lg \sin|b|$ is the Galactic 
height modulus and $M_V$ is the absolute $V$ magnitude inferred from 
the color. The left and right panels are derived using the 
solar-neighborhood CMR (1) and the $z$-dependent CMR (2), respectively. 
Crosses and circles represent stars from the 148 WFC2 fields and 162 PC1 
fields, respectively. Diagonal lines denote the $I$-band range of 
sensitivity of the fields: half of the fields have limits 
$I_{\rm max} + 5\lg \sin|b|$ that lie between the two upper-right lines, 
while one quarter lie above the upper line of these two upper-right lines 
and one quarter below the lower line. Similarly for 
$I_{\rm min} + 5\lg \sin|b|$ and the two lower-left lines. Stars 
satisfying our selection criteria (see text) fall into the large box. 
The small box is believed to contain only spheroid subdwarfs.} 
\end{figure}

The previous 53 WFC2 fields and 162 PC1 fields are described in Paper I
and Paper II in detail. Table 1 lists the characteristics of the 95 newly 
incorporated WFC2 fields in order of ascending Galactic latitude, where 
$I_{\rm max}$ and $I_{\rm min}$ are the detection threshold and the 
saturation threshold in $I$ band, respectively. The fraction of the 
4.4 arcmin$^2$ effective area of the WFC2 covered by each field is 
denoted by $\Omega$. The fields with $l\approx 180^\circ$ and 
$b\approx -22^\circ$ are near the Hyades cluster. However, by checking 
the color-magnitude diagram (CMD) of these fields, we conclude that they 
are not contaminated by cluster stars. The field near  $l\approx 355^\circ$ 
and $b\approx 23^\circ$ has one 20 s exposure and two 200 s exposures 
with the F814W filter so that the stars selected from this field appear 
substantially brighter than most WFC2 field and so make a significant  
contribution to the lower-left part of the $z-M_V$ plane (Fig.\ 1). 

\input tab1.tex

In Paper I and Paper II, the transformation from HST instrumental 
magnitudes to the standard Johnson-Cousins system was based on a 
synthetic calibration by Bahcall et al. (1994). The red end of this 
calibration was based on M giants rather than M dwarfs due to 
availability. Strictly speaking, their transformation applies to 
observations made before the WFPC2 CCD cooldown on 23 Apr.\ 1994. 
In this paper we adopt a modified form of the transformation based on an 
empirical calibration that uses M dwarfs rather than giants. This 
transformation can apply to observations made either before or after 
the cooldown by using different zero points. The difference between 
this calibration and the calibration by Bahcall et al. (1994) without 
chip-to-chip offsets is small. For example at $V-I=3$, the differences 
are: 0.02 mag in $V$ and 0.03 mag in $I$ before the cooldown; and 
0.06 mag in $V$ and 0.04 mag in $I$ after the cooldown. The difference 
between this calibration and that given by Holtzman et al. (1995) for 
the post-cooldown period is also very small: at $V-I=3$, the differences 
in $I$ and in $V$ are about 0.02 mag and 0.05 mag, respectively. We 
describe the calibration in detail in the Appendix. 
  
For each field, the extinction $A_B$ is derived from Burstein \& Heiles 
(1982). As in Paper I, we adopt $A_V=0.75A_B$, 
$A_I=0.57A_V$, $A_{V^\prime}=0.91A_V$ ($V^\prime$ denotes F606W filter) 
and $A_{I^\prime}=0.59A_V$ ($I^\prime$ denotes F814W filter) to deredden 
all the stars. 

\subsection{Sample Selection}

To derive the absolute magnitude, as in Paper I and Paper II, at first 
we adopt the solar-neighborhood CMR determined by Reid (1991):
$$ M_V=2.89+3.37(V-I) \eqno(1)$$
with a dispersion of 0.44 mag. 

The large sample substantially reduces the statistical errors relative to 
those reported in Paper II. In particular, the error in the disk scale 
length is reduced dramatically. Hence, systematic errors become more 
significant. We therefore explore the effect of metallicity variations 
on our result. As the Galactic height increases, more and more disk 
dwarfs with lower metallicity will be detected, and these tend to be less 
luminous than the dwarf stars near the Galactic plane at the same $V-I$ 
color. Hence we modify the above CMR by adding a term that varies with 
the Galactic height $z$:
$$ M_V=2.89+3.37(V-I)+f(z)m(V-I) , \eqno(2)$$
where 
$$
f(z)= \left\{\begin{array}{ll}
             \frac{|z|}{1.5\,{\rm kpc}},&  |z|\leq1.5\,{\rm kpc} \\
             1,                         &   |z|>1.5\,{\rm kpc} 
             \end{array}
      \right.
$$              
and
$$
m(V-I)=\left\{\begin{array}{ll}
             0.2(V-I),        &  (V-I)\leq2.5 \\
             1-0.2(V-I),      &  (V-I)>2.5 
             \end{array}
      \right. .
\eqno(3)
$$
Figure 1 in Reid (1991) and Figure 10 in Monet et al. (1992) show a 
color-dependent dispersion in the color-magnitude relation: the 
dispersion is peaked at $(V-I) \sim 2.5$ and becomes smaller towards 
redder and bluer colors. We estimate this dispersion to be $m(V-I)$ 
as given in equation (3). We then also adopt this $m(V-I)$ as the scale of 
the offset in $M_V$ as a function of height (eq.\ [2]). That is, we assume
that the observed scatter is due to metallicity variation and hence assume 
that the amplitude of the metallicity effect with height is proportional to 
this scatter. It should be pointed out that this {\it ad hoc} CMR does not 
necessarily represent the true metallicity effect on main sequence stars at 
different Galactic height. However, this CMR is adequate for our purposes:
making a first-order correction for the metallicity effect and estimating 
the systematic errors due to this correction.
 
Stars in our sample are chosen to satisfy both a luminosity criterion and 
a Galactic height criterion: $ 8.0 < M_V < 18.5 $. This corresponds to 
$ 1.53 < V-I < 4.63 $ under the CMR (1). The blue boundary prevents 
contamination by spheroid giants (Green, Demarque, \& King 1987), and the 
red boundary is about the red edge of the color-magnitude diagram of Monet 
et al.\ (1992) (although the CMR becomes double-valued at $V-I \sim 4.4$ in 
this diagram, the relatively small number of dwarfs fainter than $M_V=18.5$ 
makes this effect negligible in our analysis); Galactic height $z$ must be 
below 3200 pc if the solar-neighborhood CMR (1) is used and below 2400 pc 
if the $z$-dependent CMR  (2) is used in order to avoid the contamination by 
spheroid dwarfs (see Paper I \& Paper II). 

When CMR (1) is adopted, altogether, in the 148 WFC fields and 162 PC1 
fields, 1413 stars satisfy our selection criteria: 263 in the fields 
analyzed in Paper I, 85 in the 31 additional fields analyzed in Paper II, 
and 1065 in the 95 new fields. Note that the numbers of stars in the 
previous fields differ slightly from those in Paper I and Paper II 
because of a slight change in the transformations from WFPC2 instrumental 
magnitudes to standard Johnson-Cousins magnitudes as mentioned in \S\ 2.1.
If we use the modified CMR (2), the total number of selected stars is 1373.

The distribution of stars in the sample as a function of $R$, the 
cylindrical distance from the Galactic center and as a function of 
height, $z$, above the plane is shown in Figure 2 for the cases based 
on the solar-neighborhood CMR (1) and the $z$-dependent CMR (2). 
The mean and standard deviation of Galactocentric radius of all stars is 
$\overline{R}=7.7 \pm 1.7$ kpc for CMR (1) and 
$\overline{R}=7.7\pm 1.4$ kpc for CMR (2). The lower panels of Figure 2 
show the histogram of the heights above the Galactic plane for each case.
We plot the fraction of stars weighted by $(R-\overline{R})^2$ as well 
as the nonweighted one. The weight factor comes from equation (2.1) 
and (2.3) of Gould (1995) with $f(R;H)=\exp(-R/H)$. The weighted plot 
tells us the stars at which height dominate our derivation of the slope of 
the radial distribution of disk stars and thus of the scale length of 
the disk. There is almost no weight from stars with height $z <1$ kpc. 
The average weighted height $\overline{z}_*$ is about 2 kpc for CMR (1) 
and is about 1.5 kpc for CMR (2). As we discuss in \S\ 3, this implies 
that our final results on the disk profiles are primarily based on 
stars well above the thin disk population.

\begin{figure}
\centerline{\psfig{figure=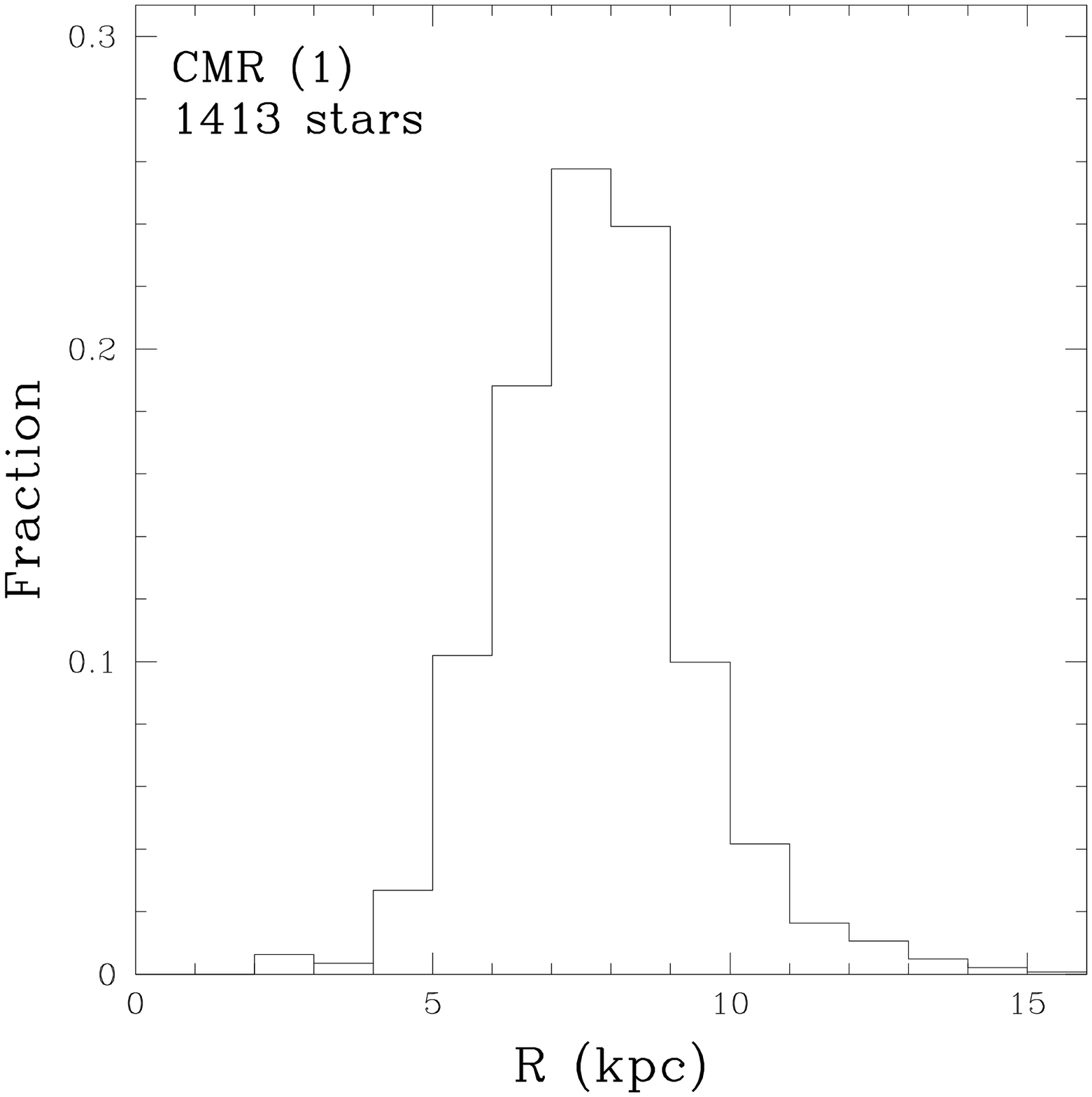,height=8cm,width=8cm}
            \psfig{figure=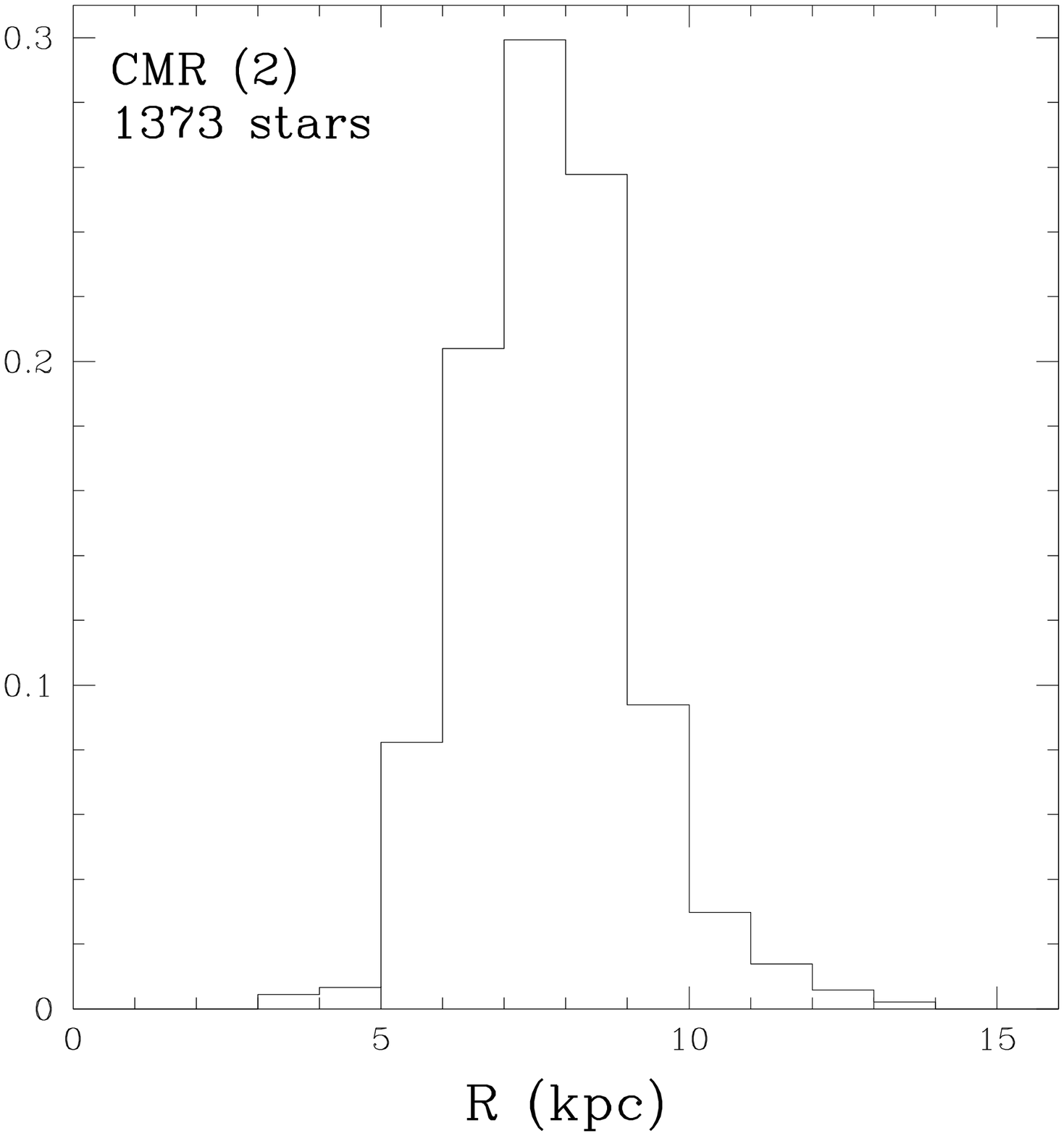,height=8cm,width=8cm}}
\centerline{\psfig{figure=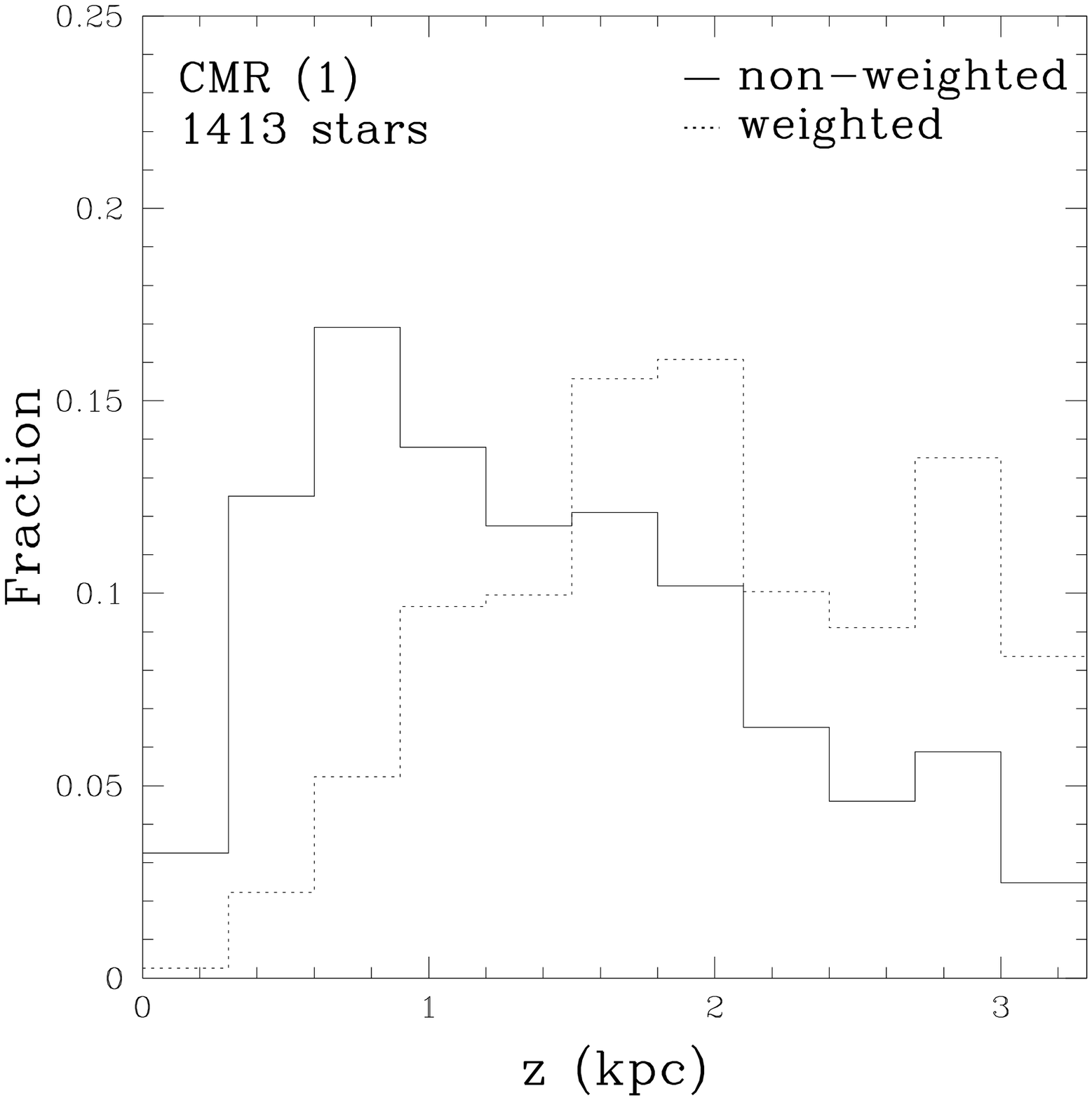,height=8cm,width=8cm}
            \psfig{figure=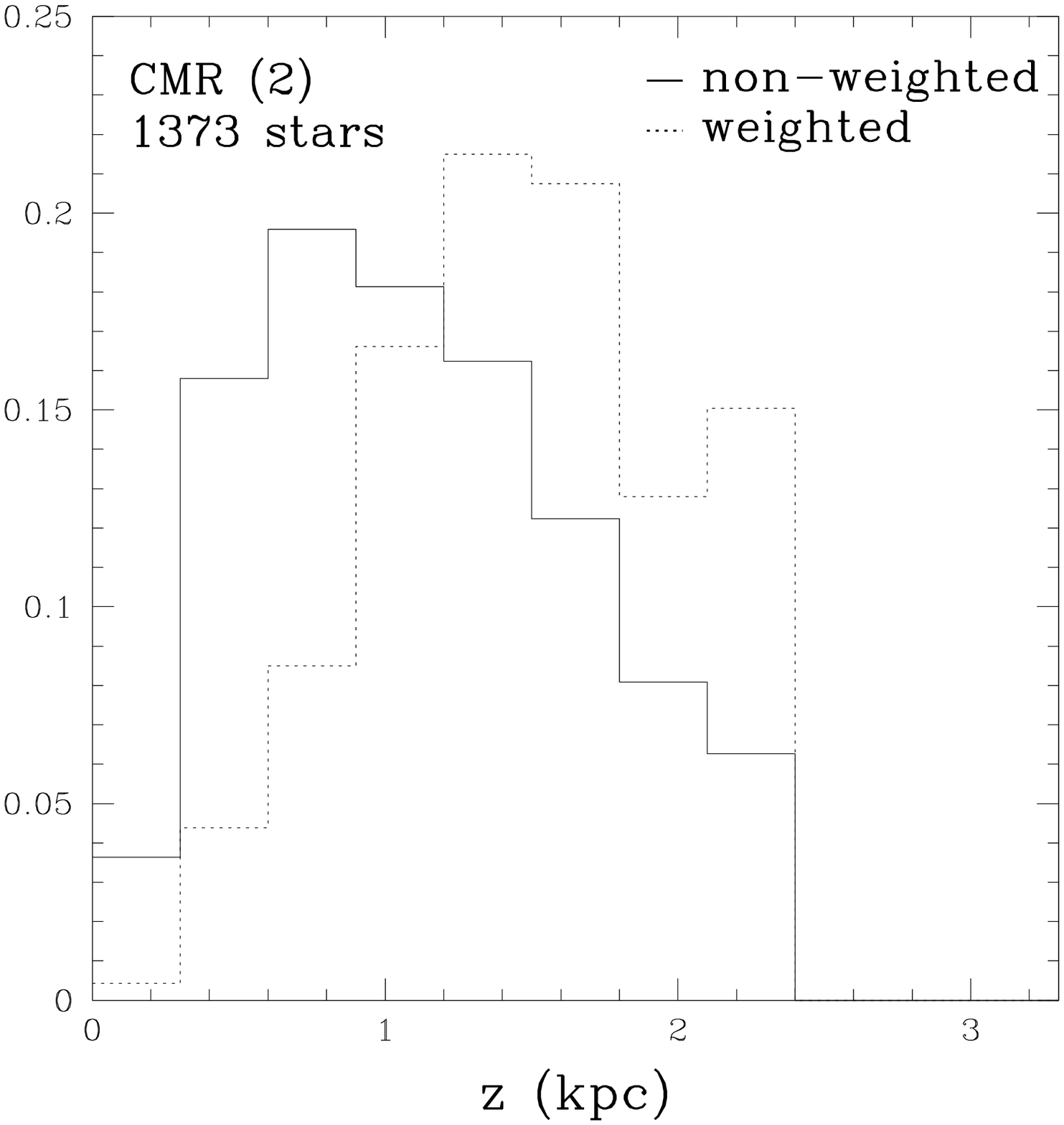,height=8cm,width=8cm}}

\caption[]{The distribution of stars in our sample as a function of the
Galactic coordinates ($R$,$z$). CMR (1) and (2) are 
used in the left and right histograms, respectively. The Galactic height 
distribution weighted by $(R-\overline{R})^2$ is shown as dotted lines in 
the lower panels.}
\end{figure}

\subsection{Models and Method}

The luminosity function (LF) of disk M dwarfs is modeled as a function of 
Galactic position ($R$, $z$) by 
$$\Phi(i,z,R)=\Phi_i\nu(z)\exp\biggl(-{R-R_0 \over H}\biggr), \eqno(4)$$
where $\Phi_i$ is the LF for the $i$th magnitude bin in the solar 
neighborhood, $R_0= 8$ kpc is the Galactocentric distance of the Sun, 
and $H$ is the scale length of the disk. The density profile
$\nu(z)$ is assumed to have either a ``sech$^2$" form,  
$$\nu_s(z)=(1-\beta){\rm sech}^2 {z \over h_1} +\beta \exp 
\biggl(-{|z| \over h_2}\biggr) \eqno(5)$$ 
or a ``double exponential" form,
$$\nu_e(z)=(1-\beta) \exp\biggl(-{|z| \over h_1}\biggr) +\beta \exp 
\biggl(-{|z| \over h_2}\biggr). \eqno(6)$$ 

The method of maximum likelihood (see Paper I) is applied to derive 
simultaneously the LF $\Phi_i$ at each magnitude bin and the disk 
profile parameters ($h_1$, $h_2$, $\beta$). The magnitude bins are 
centered at $M_V=$ 8.25, 9, 10, 11, 12, 13, 14, 15.5, 17.5, respectively. 
The size of each bin is 1 mag except the first one (0.5 mag) and the 
last two (2 mag). Given a value of the disk scale length $H$, a solution 
of the above 12 parameters is found by maximum likelihood. Then, the 
scale length is determined by maximizing the likelihood over the 
ensemble of solutions using different values of $H$.

\section{RESULTS}

The derived parameters fall into two categories: the LF $\Phi_i$ at each 
magnitude bin and the disk profile parameters such as the scale length. 
As discussed in Paper I and Paper II, there are almost no stars near 
the Galactic plane in our sample so that we lack information on the 
local stellar density. Hence, there is a degeneracy between the best 
fits of the sech$^2$ model and the double exponential model which cannot
be resolved by our HST data alone. Therefore we must normalize the HST 
LF using LFs derived by other methods. We therefore discuss the LF first 
and then examine the disk parameters. Finally we convert the LF to a 
mass function.   

\subsection{Luminosity Function}

The best fit LF for the sech$^2$ model and that for the double 
exponential model have nearly the same shape but differ from each other 
in normalization. For example, in the case of using the $z$-dependent 
CMR (2), the exponential model has a normalization 1.5 times that of 
the sech$^2$ model. This normalization difference is compensated by the 
vertical density profile: in the same example, the vertical density of 
the sech$^2$ model is almost 1.5 times larger than that of the 
exponential model everywhere except in the vicinity of the plane 
(where there are few data). We therefore perform a linear combination 
of the LFs in these two models. The combination coefficient is obtained 
by normalizing the combined LF using the local LF in the region 
$8.5 \leq M_V \leq 12.5 $ derived by Wielen, Jahreiss, \& Kr\"uger (1983). 
It turns out that the sech$^2$ model agrees fairly well with the 
local-star normalization: for CMR (1), the relative difference from the 
local normalization is less than 0.3\%; for CMR (2), the relative 
difference from the local normalization is less than 10\%. In the case of 
the double exponential model, the relative differences from the local 
normalization are much larger, 93\% and 65\% for CMR (1) and CMR (2), 
respectively. Note that the relative errors of the LF given by 
Wielen et al.\ (1983) increase from $\sim$10\% to $\sim$30\% in the 
above magnitude range.  

\begin{figure}
\centerline{\psfig{figure=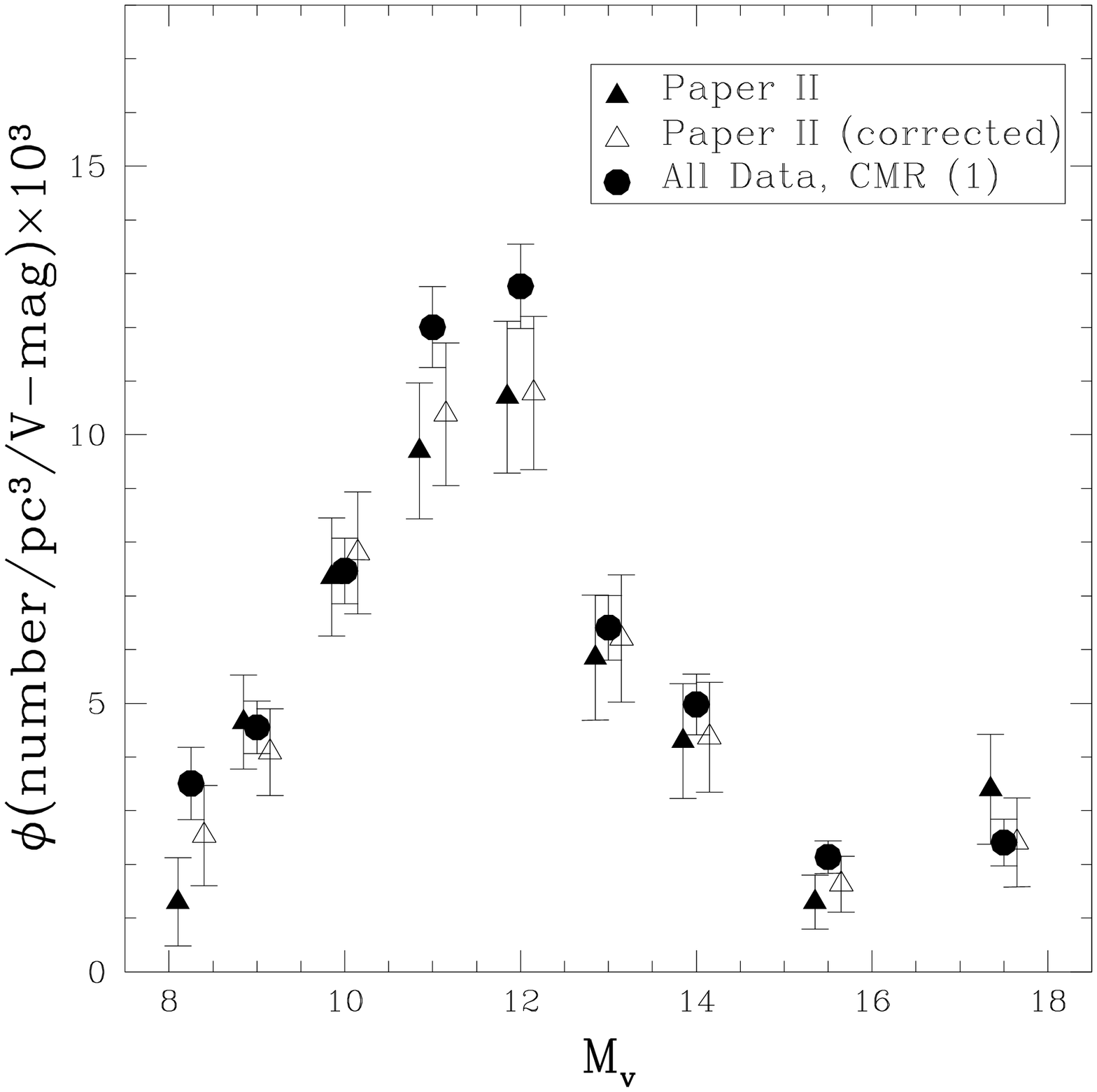,height=8cm,width=8cm}
            \psfig{figure=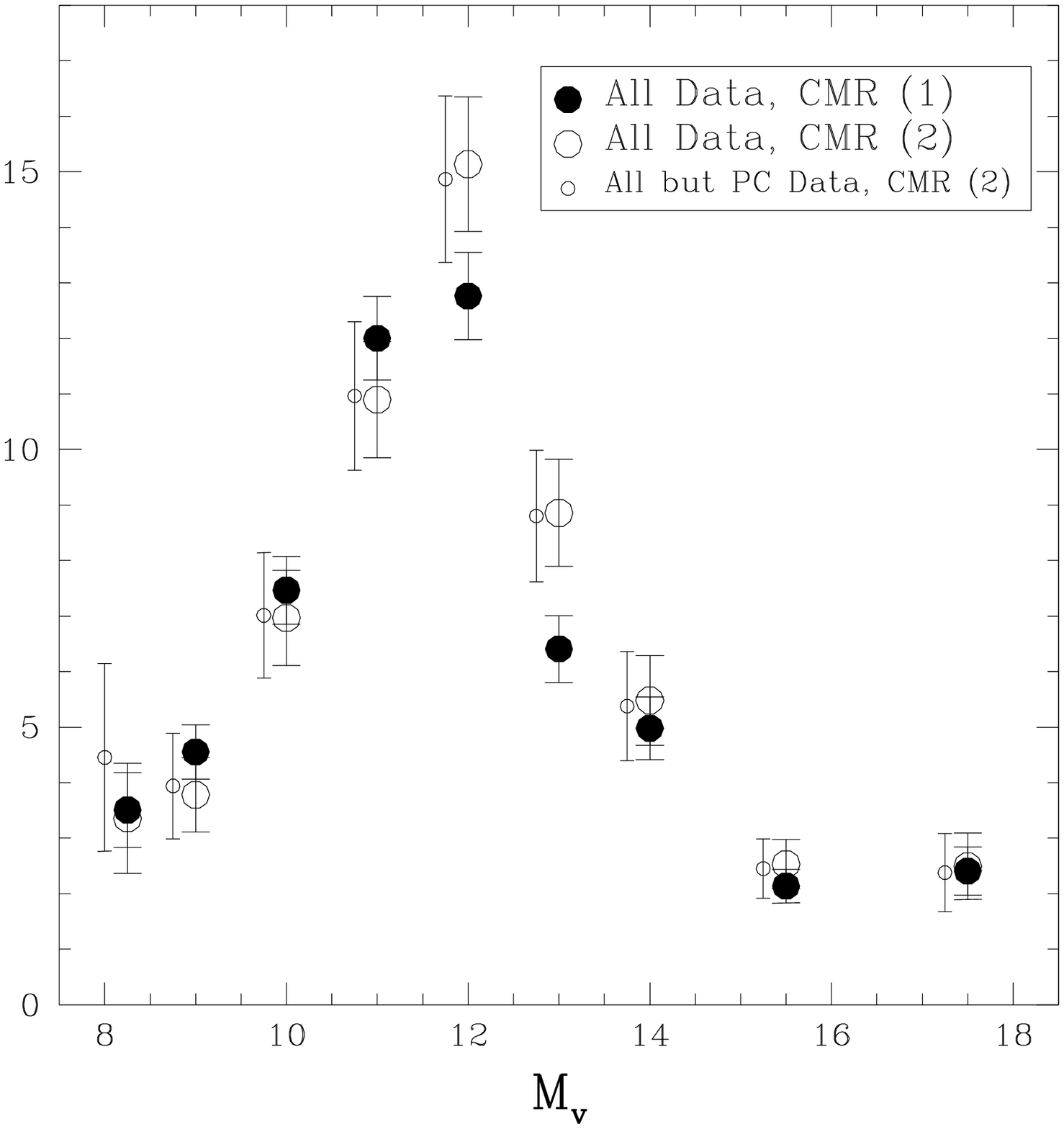,height=8cm,width=8cm}}
\caption[]{Normalized luminosity functions derived using the method of 
maximum likelihood. The left panel shows the 
comparison of LFs determined using different data sets. Here the 
solar-neighborhood CMR (1) is adopted. The result of Paper II and that 
of Paper II with the correction of error estimation and magnitude 
transformation are displayed together with the LF determined by all the 
data available. These three LFs are slightly displaced along the 
horizontal direction with respect to each other in order to make the 
figure easier to read. The right panel shows the comparison between
LFs determined using CMR (1) and the $z$-dependent CMR (2). For 
comparison, in the right panel we also plot the LF (slightly displaced 
along the horizontal direction) derived using only WFC2 data and 
CMR (2) (see the text for details).}
\end{figure}

\begin{deluxetable}{rrr}
\tablecaption{Normalized Luminosity Functions
              \label{tab:LF}}
\tablehead{M$_V$ & $\phi$ [CMR(1)] & $\phi$ [CMR(2)] \\
(mag)& (10$^{-3}$pc$^{-3}$V-mag$^{-1}$) & (10$^{-3}$pc$^{-3}$V-mag$^{-1}$)}
\startdata
 8.25  &  3.51$\pm$0.67  &    3.36$\pm$0.99  \\
 9.00  &  4.56$\pm$0.49  &    3.78$\pm$0.67  \\
10.00  &  7.46$\pm$0.61  &    6.97$\pm$0.86  \\
11.00  & 12.00$\pm$0.75  &   10.90$\pm$1.05  \\
12.00  & 12.76$\pm$0.78  &   15.14$\pm$1.21  \\
13.00  &  6.41$\pm$0.60  &    8.86$\pm$0.96  \\
14.00  &  4.98$\pm$0.57  &    5.48$\pm$0.81  \\
15.50  &  2.14$\pm$0.30  &    2.53$\pm$0.44  \\
17.50  &  2.41$\pm$0.43  &    2.49$\pm$0.60  \\
\enddata
\bigskip
\begin{footnotesize}
(Both LFs are derived using all the available data.)
\end{footnotesize}
\end{deluxetable}

We compare the LFs derived using different data sets in Figure 3. 
In Table 2, we also list LFs derived using all the available data 
and adopting different CMRs.
Assuming CMR (1), our new result is consistent with those derived in 
Paper I and Paper II. The statistical errors are reduced considerably. 
The last data point in this plot (centered at $M_V=17.5$ with a bin 
width of 2 magnitudes) drops from $3.4 \times 10^{-3} 
{\rm pc}^{-3} {\rm mag}^{-1}$ to $2.5 \times 10^{-3} {\rm pc}^{-3} 
{\rm mag}^{-1}$. The main reason for this is our correction of the error
estimation as well a small contribution from the new transformations 
that we adopt. This last data point had caused a worry in Paper I and 
Paper II because it was much higher than the result from naive binning. 
The new result confirms that the maximum likelihood method yields a 
reasonable value for the last point. 
We note that the good agreement between the Paper II results based on 
naive binning and the corrected maximum-likelihood calculation shows that 
the effects of Malmquist bias are small.  This is because naive binning 
ignores the dispersion in the CMR, and so ignores Malmquist bias, 
while maximum likelihood automatically compensates for Malmquist bias 
(assuming that the adopted dispersion in the CMR is correct).  
Ignoring the first half-magnitude bin (for which boundary
effects are important) the fractional difference in the two methods 
averages $\sim 10\%$.  As we argued in Papers I and II, the small
size of this effect is due to the fact that the sample
extends to the ``top" of the disk.

The LF peaks at about $M_V=12$ and still shows a sharp drop towards 
$M_V=14$, as mentioned in Paper I and Paper II. Reid \& Gizis (1997) 
showed that there might be some fine structures in the CMR (see their 
Fig. 13). We attempted to use their new analytic fit of the CMR in 
our analysis. It leads to a much steeper slope of the LF in the range of 
12 mag to 14 mag. The physical reason is that their fit has a relatively 
larger slope, 9.74 instead of 3.37, in this region. This larger slope 
makes the magnitude range $12 < M_V < 14$ correspond to a smaller color 
range ($\sim 0.2$ mag) and hence fewer stars. However, it is possible that 
the ``steep" part of the CMR discussed by Reid \& Gizis (1997) may differ 
for lower-metallicity stars (most likely shifted blueward, see Fig. 7 of Gizis 
1997). Such a blueward shift in the steep section of the CMR would help 
smooth the LF, and would be quite plausible because the
majority of the HST sample is likely to be more metal poor than the
Reid \& Gizis (1997) stars.  However, there do not appear to be any
available data sets at lower metallicity that would allow us to
verify the reality of such a shift nor to measure its size.  We therefore
adopt the original Reid (1991) law (see eq. (1)).  We note, however, that 
this means that we cannot be sure that the sharpness of the drop in the LF 
between $M_V=12$ and $M_V=14$ is real.

Adopting the $z$-dependent CMR (2) leads to the LF having a slight 
horizontal shift towards the faint end with respect to the above LF. 
This is not surprising since at the same color, stars tends to be 
fainter under this CMR. The LF still peaks at $M_V=12$ with a higher 
peak value and becomes more symmetric about this peak. The biggest 
difference in LFs for cases of CMR (1) and CMR (2) occurs around the 
peak. This is the reflection of the fact that the $z$-dependent
CMR (2) has, at a given Galactic height, the largest correction at 
$V-I=2.5$ and smaller corrections toward redder and bluer colors. 

The consistency of the results presented here with those derived in 
Paper I and Paper II (the latter based on only about 1/4 of the final 
sample) demonstrates the stability of our method and also of the 
underlying data set which has been assembled from 
{\it Hubble Space Telescope} observations over almost a decade.
Because these data were taken by two different instruments (WFC2 and PC1),
it is also important to compare the LF found using all the data with
that derived using only WFC2 data. Since the PC1 data comprise
only about 5\% of the full sample, one might not expect their exclusion 
to have much impact.  However, as was pointed out in the discussion of 
Figure 1, the PC1 data tend to dominate the sample for the bright end of 
the LF.  We see from Figure 3 that indeed inclusion of these data reduces 
the error bars of the first few LF bins by 25--40\%.  Note that in the 
brightest bin where the PC1 data clearly dominate, the two LFs are 
consistent. We conclude that the underlying data set is self-consistent.

\subsection{Disk Parameters}

We summarize the best-fit disk parameters for M dwarfs in Table 3. 
The first two lines are the original results in Paper I and Paper II, 
respectively. The third line is the corrected result for Paper II: 
the error estimation mentioned in \S\ 1 is corrected and the new 
transformation formulae are applied. Next are the best fits using 
only WFC2 data with the solar-neighborhood CMR (1) and the 
$z$-dependent CMR (2) being adopted, respectively. Finally, we list 
the best fits using all the data available with CMR (1) and CMR (2), 
respectively. For the last four cases, we list the results of the 
sech$^2$ model, the double exponential model and the normalized model 
(see \S\ 3.1). To derive the local mass density $\rho_0$ and the 
column density $\Sigma_M$ of M dwarfs, we adopt the mass-luminosity 
relation given by Henry \& McCarthy (1993) for CMR (1) and an 
interpolation between relations based on different metallicities 
(Baraffe et al.\ 1998) for CMR (2) (see \S\ 3.3 for details).

\begin{deluxetable}{lcccccc}
\tablecaption{Best-Fit  Models for M Stars ($8<M_V<18.5$)
              \label{tab:fit}}
\tablehead{Data Set & $h_1$ & $h_2$ & $\beta$ & $\rho_0$ & $\Sigma_M$ & $H$ \\            & (pc)   & (pc)  & (\%)    & ($M_\odot$pc$^{-3}$) & ($M_\odot$pc$^{-2}$) & (kpc)}
\startdata
Paper I   & 323$\pm$54 & 656$\pm$78 &  19.8$\pm$7.1 & 0.0159$\pm$0.0044 & 12.4$\pm$1.9 & 3.02$\pm$0.43  \\
Paper II  & 320$\pm$50 & 643$\pm$60 &  21.6$\pm$6.8 & 0.0158$\pm$0.0041 & 12.3$\pm$1.8 & 2.92$\pm$0.40  \\
Paper II (corrected)   & 314$\pm$52 & 627$\pm$55 &  23.7$\pm$7.3 & 0.0161$\pm$0.0043 & 12.5$\pm$1.8 & 2.95$\pm$0.41  \\
\hline
All but PC Data, CMR(1) & & & & &  \\
sech$^2$ Model & 327$\pm$32 & 604$\pm$28 &  21.9$\pm$4.2 & 0.0207$\pm$0.0040 & 16.1$\pm$1.9 & 3.28  \\
exp Model & 185$\pm$21 & 609$\pm$30 & 10.1$\pm$2.3 & 0.0437$\pm$0.0106 & 19.9$\pm$2.7 & 3.28  \\
normalized & - & - & - & 0.0183 & 15.7$\pm$2.1 & 3.28  \\
\hline
All but PC Data, CMR(2) & & & & &  \\
sech$^2$ Model & 264$\pm$60 & 435$\pm$18 &  53.6$\pm$13.0 & 0.0191$\pm$0.0049 & 13.6$\pm$1.8 & 2.75  \\
exp Model & 149$\pm$40 & 434$\pm$19 &  33.2$\pm$13.2 & 0.0309$\pm$0.0125 & 15.1$\pm$2.6 & 2.75  \\
normalized & - & - & - & 0.0156 & 13.1$\pm$2.5 & 2.75  \\
\hline
All Data, CMR(1) & & & & &  \\
sech$^2$ Model & 332$\pm$31 & 609$\pm$28 &  24.4$\pm$4.2 & 0.0179$\pm$0.0030 & 14.3$\pm$1.4 & 3.28$\pm$0.18  \\
exp Model & 193$\pm$22 & 614$\pm$31 & 12.1$\pm$2.5 & 0.0347$\pm$0.0071 & 16.9$\pm$1.8 & 3.28$\pm$0.18  \\
normalized & - & - & - & 0.0180 & 14.3$\pm$1.3 & 3.28$\pm$0.18  \\
\hline
All Data, CMR(2) & & & & &  \\
sech$^2$ Model & 270$\pm$55 & 440$\pm$18 &  56.5$\pm$11.5 & 0.0169$\pm$0.0034 & 12.4$\pm$1.3 & 2.75$\pm$0.16  \\
exp Model & 156$\pm$40 & 439$\pm$19 &  38.1$\pm$12.6 & 0.0253$\pm$0.0081 & 13.4$\pm$1.8 & 2.75$\pm$0.16  \\
normalized & - & - & - & 0.0153 & 12.2$\pm$1.6 & 2.75$\pm$0.16  \\

\enddata
\bigskip
\begin{footnotesize}
(Only best-fit sech$^2$ models are listed here for Paper I and Paper II.)
\end{footnotesize}

\end{deluxetable}

The statistical uncertainties are reduced considerably when all the 
available data are taken into account. The new results are basically 
consistent with the previous ones. Stars in the sample are detected as 
far as on the ``top" of the disk. See Figure 2. The average height for 
the weighted distribution is about 2 kpc for the solar-neighborhood 
CMR (1). Therefore, the scale length $H$ we measure is in fact that 
of the kinematically hottest and hence most metal-poor component of 
the disk, which is often called the ``old" or ``thick" disk. Optical
studies prior to 1990 support a disk scale length of 3.5-4.5 kpc (see 
Sackett 1997 and references therein). More recent studies, on the 
contrary, tend to give a shorter scale length. Paper I and Paper II 
give estimates of the scale length to better precision than other methods. 
Our results here also favor a short scale length, 3.3 kpc. This is the most 
precise measurement of the scale length of the old or thick disk up to now 
with a statistical uncertainty less than 6\%. At this stage, the systematic 
errors become significant with respect to the statistical errors which 
motivates our modification of the CMR. If the $z$-dependent CMR (2) is 
adopted, stars tend to be fainter and therefore closer at a given color, 
and the disk scale length is therefore smaller, i.e. $2.75$ kpc. We 
estimate the systematic error to be half of the difference between the 
best-fit results using the solar-neighborhood CMR (1) and $z$-dependent 
CMR (2), respectively. We finally express the disk scale length to be 
$H = 2.75 \pm 0.16 {\rm (statistical)} \pm 0.25 {\rm (systematic)}$ kpc.
Our measurement is based on low mass stars which dominate the stellar mass 
in the disk so that the scale length we derive is likely to reflect the 
stellar mass distribution in the disk. 

\subsection{Mass Function}

\begin{figure}
\centerline{\psfig{figure=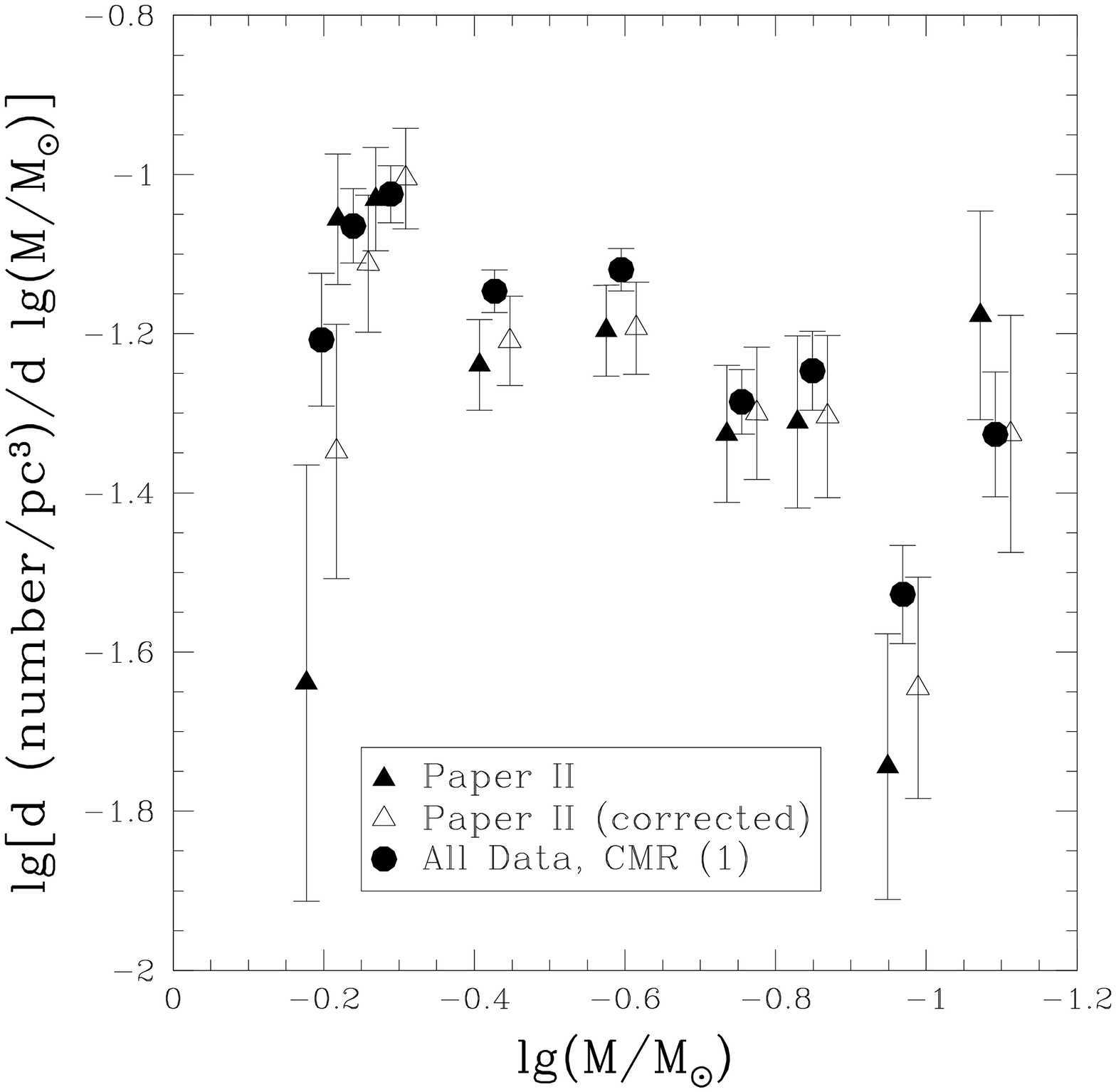,height=8cm,width=8cm}
            \psfig{figure=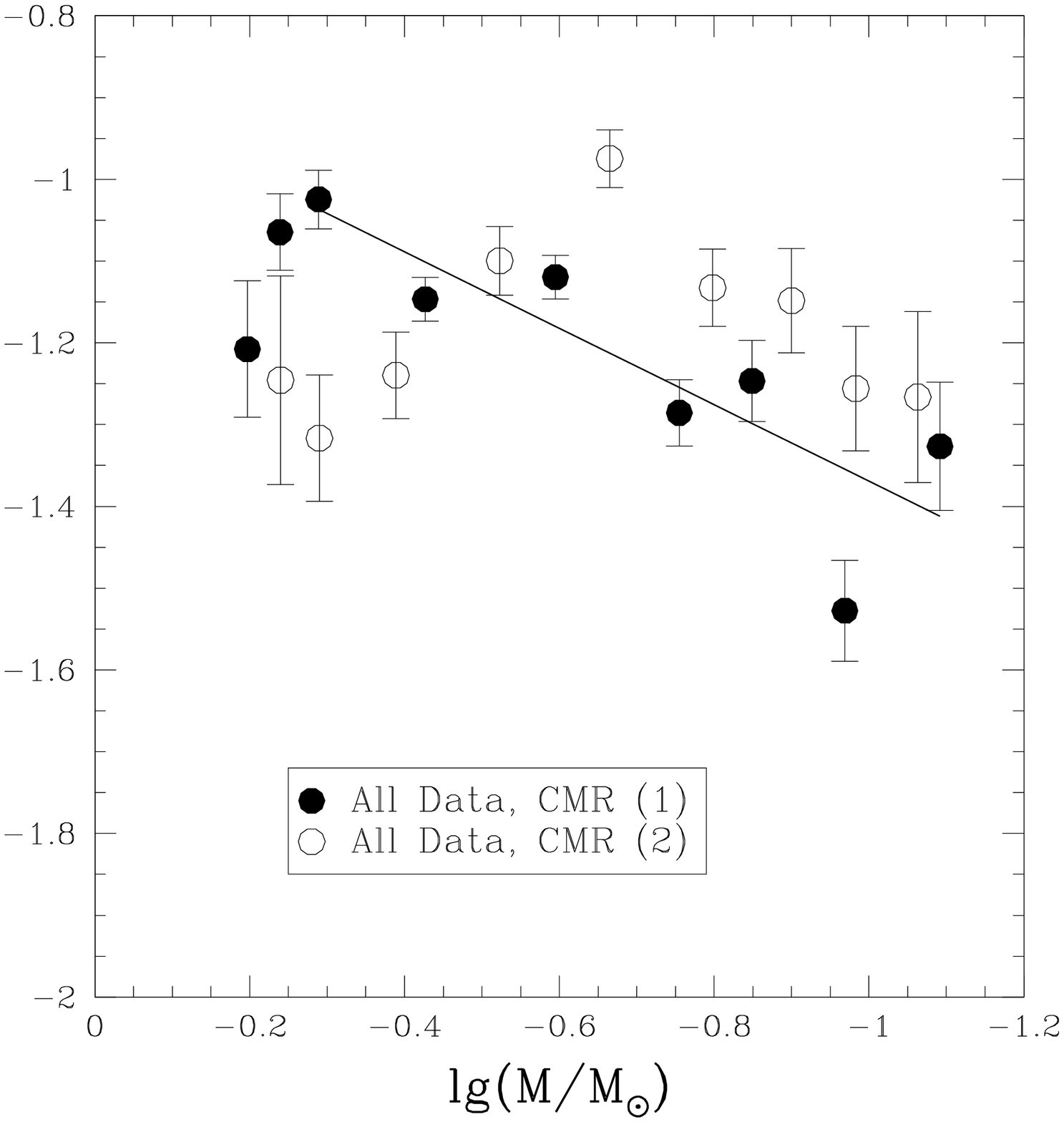,height=8cm,width=8cm}}
\caption[]{Mass functions from LFs. The symbols and labels have the same
meanings as those in Fig.\ 3. In the left panel, the three MFs are slightly
displaced along the horizontal direction with respect to each other. 
In the right panel, we also show the best 
power law fit to the MF in the case of using the solar-neighborhood 
CMR (1): $\lg\Phi_M=-0.90+0.47 \lg(M/M_\odot)$ (the first two data points 
are excluded from the fit), where $\Phi_M\equiv dN/d \lg M$ and $N$ is 
the number density of stars.}
\end{figure}

\begin{deluxetable}{cccc}
\tablecaption{Normalized Mass Functions
              \label{tab:MF}}
\tablehead{CMR(1) & & CMR(2) & \\
$\lg(M/M_\odot)$ & $\lg\Phi_M$ & $\lg(M/M_\odot)$ & $\lg\Phi_M$}
\startdata
-0.20 & -1.21$\pm$0.08 & -0.24 & -1.25$\pm$0.13 \\
-0.24 & -1.06$\pm$0.05 & -0.29 & -1.32$\pm$0.08 \\
-0.29 & -1.02$\pm$0.04 & -0.39 & -1.24$\pm$0.05 \\
-0.43 & -1.15$\pm$0.03 & -0.52 & -1.10$\pm$0.04 \\
-0.60 & -1.12$\pm$0.03 & -0.67 & -0.97$\pm$0.04 \\
-0.76 & -1.29$\pm$0.04 & -0.80 & -1.13$\pm$0.05 \\
-0.85 & -1.25$\pm$0.05 & -0.90 & -1.15$\pm$0.06 \\
-0.97 & -1.53$\pm$0.06 & -0.98 & -1.26$\pm$0.08 \\
-1.09 & -1.33$\pm$0.08 & -1.06 & -1.27$\pm$0.10 \\
\enddata
\bigskip
\begin{footnotesize}
(Both MFs are derived using all the available data. Here $\Phi_M$ is 
defined as $d {\rm N(number/pc^3)}/d \lg M$.)
\end{footnotesize}
\end{deluxetable}

To convert the LF to the mass function (MF) for the solar-neighborhood 
CMR (1), we use the empirical mass-$M_V$ relation given by 
Henry \& McCarthy (1993) (see their eq. [5]). The mass function 
$\Phi_M$ is defined as the number density of stars $N$ per decade in 
mass and it usually can be characterized by a power law:
$$\lg \Phi_M\equiv \lg(dN/d \lg M)= \alpha\lg(M/M_\odot) + \beta. \eqno(7)$$ 

Similar to the results of Papers I \& II, the MF in the range of 
$ 0.08 M_\odot < M < 0.6 M_\odot$ obtained in this paper tends to
fall towards the lower mass (see Fig.\ 4). The statistical errors are 
now reduced considerably, especially for the first and the last two bins. 

There are two structures in the MF derived from CMR (1). First, there 
appears to be a break in the slope of the MF at about $0.5M_\odot$. 
The MF tends to rise from $\sim 0.6 M_\odot$ ($\lg(M/M_\odot) =-0.2$) 
to $\sim 0.5M_\odot$ ($\lg(M/M_\odot)=-0.3$) and after that it drops 
toward the lower mass.  Whether there is a break in the MF still is a 
matter of debate. The MF from the LF of Wielen et al. (1983) shows 
such a break. The results of Papers I and II, combined with the Wielen et 
al. (1983) MF at higher masses, reinforce the case for a break 
(see Fig.\ 3 of Paper II). Taking Hipparcos data into account changes 
somewhat the Wielen et al. (1983) LF.\  See the 25 pc limited sample of 
Jahreiss \& Wielen (1997). These changes cause the first four points in 
Figure 3 of Paper II to increase somewhat, but the break in the slope 
is still pronounced. However, Reid \& Gizis (1997) analyzed an 8-pc 
sample and found no evidence for a break point in the MF. Instead the MF 
rises gradually toward the lower mass (see their Fig.\ 4). The hint of 
a break point in the MF we derived for CMR (1) is at the level of
$2\sigma$, so it is very marginal. The second structure in the MF
is the dip at $\sim 0.1M_\odot$. This dip is detected at the 
$3\sigma$ level and so is of high statistical significance. 
However, it is possible that this dip is an artifact of subtleties 
in the structure of the CMR or mass-luminosity relation that are not 
reflected in our simple model.
 
For the solar-neighborhood CMR (1), the power law fit to the MF 
agrees very well with the result in Paper II. After correcting the error 
estimation and applying the new photometry transformation for data used 
in Paper II, we get a power law fit of the MF with a slope 
$\alpha=0.43 \pm 0.11$ and a zeropoint $\beta=-0.97$. When all the 
available data are taken into account, the slope becomes 
$\alpha=0.41 \pm 0.06$ and $\beta=-0.94$. If we do the fit only using 
the data points after the possible break point,
$M \sim 0.5 M_\odot$, we get $\alpha=0.47 \pm 0.07$ and $\beta=-0.90$.

For CMR (2), we modify the CMR to reflect the metallicity gradient above 
the Galactic plane. Therefore, to derive the MF in this case, the 
metallicity effect on the mass-$M_V$ relation must also be taken into 
account. Less luminous stars are most likely to be nearby stars 
(see Fig.\ 1) and are therefore little affected by the metallicity 
gradient. We assume that, from the plane to a height $z=1500$ pc, the 
metallicity changes linearly from [M/H]$=0.0$ to [M/H]=$-0.5$ and keeps 
the value $-0.5$ above $z=1500$ pc. We calculate the mean height above 
the Galactic plane of the stars in our sample in each magnitude bin. We 
then obtain a linearly interpolated mass-$M_V$ relation at these mean 
heights, respectively, using mass-$M_V$ relations for [M/H]=$0.0$ and 
[M/H]=$-0.5$ given by Baraffe et al.\ 1998. The [M/H]=$0.0$ mass-$M_V$ 
relation by Baraffe et al.\ 1998 agrees with that by Henry \& 
McCarthy (1993).  The resultant MF is shown in Fig.\ 4. This MF has a dip 
at $\sim 0.5 M_\odot$ and a peak at $\sim 0.2 M_\odot$. However, these 
features may be caused by the somewhat {\it ad hoc} CMR and the 
averaged mass-$M_V$ relation that we adopt. Thus, the detailed structure 
in the MF should not be given much credence.  More reliable is the 
overall slope $\alpha=-0.10$. In Paper II, we advocated a correction to 
the slope due to unresolved binaries of $\Delta\alpha\sim -0.35$. Our 
final slope corrected for binaries is therefore 
$$ 
\alpha \sim -0.45\qquad \rm [CMR\ (2),\ corrected\ for\ binaries].
\eqno(8)
$$

MFs derived using all the available data and adopting different CMRs are 
tabulated in Table 4.

\subsection{Mass Density and Microlensing}

As we described in \S\ 3.1, the LF we finally get is the linear 
combination of LFs from the sech$^2$ model and the double exponential 
model that satisfies the local normalization (Wielen et al.\ 1983) 
in the range $8.5 \leq M_V \leq 12.5$. The normalized local mass 
densities for the solar-neighborhood CMR (1)  
is  $\rho_0 = 0.0180 M_\odot {\rm pc}^{-3}$ which is about 12\% higher 
than those in Paper I and II but is consistent at the $1\sigma$ level.
The normalized local mass densities for the $z$-dependent CMR (2) is 
$\rho_0 = 0.0153 M_\odot {\rm pc}^{-3}$.

The normalized local surface density for CMR (1) is   
$\Sigma_M \sim 14.3 M_\odot {\rm pc}^{-2}$, which is 15\% higher 
than those in Papers I \& II, while the normalized local surface 
density for CMR (2) is $\Sigma_M \sim 12.2 M_\odot {\rm pc}^{-2}$. 
Our estimate of the total column density of the disk (gas plus stars, 
see Paper I) does not change much ($\Sigma_{obs}\sim 43 M_\odot 
{\rm pc}^{-2}$ for CMR (1) and $\Sigma_{obs}\sim 41 M_\odot {\rm pc}^{-2}$ 
for CMR (2)).

How much do the stellar contents contribute to the optical depth of 
microlensing towards the Large Magellanic Cloud (LMC)? For the exponential 
component of a disk, the optical depth $\tau=2\pi G\Sigma h \csc^2b/c^2$, 
and for the sech$^2$ component, the optical depth $\tau=2 (\ln 2) 
\pi G\Sigma h \csc^2b/c^2$ where $b=-33^\circ$ is the Galactic latitude 
of the LMC and $\Sigma$ is the local surface density of stars. The 
estimate of the optical depth from our data is about $1\times 10^{-8}$ 
for either CMR (1) or CMR (2). Thus the disk stars contribute only 
$ \sim 8\%$ of the optical depth $1.2\times 10^{-7}$ determined by the 
MACHO collaboration (Alcock et al.\ 2000a). If we assume the disk profile 
we report here extends all the way to the Galactic center, we can also 
estimate the contribution of disk stars to the optical depth of 
microlensing towards sources in Baade's Window ($(l,b)=(1^\circ,-4^\circ)$).
We further assume that other stars are distributed like M stars and the 
density ratio of other stars to M dwarfs is constant. Under these
conditions, the optical depth is estimated to be $4.1\times 10^{-7}$ for CMR 
(1) or $5.2\times 10^{-7}$ for CMR (2) which is only about 20\% of 
the value measured by the MACHO collaboration (Alcock et al.\ 2000b). 
These numbers are very close to the original estimates of Paczy\'nski (1991)
and Griest et al.\ (1991). Subsequently, Kiraga \& Paczy\'nski (1994) 
realized that the optical depth toward Baade's Window is dominated by 
bulge stars, not disk stars.  However, this close agreement is something 
of a coincidence.  In this paper we derive both a lower local stellar 
density and a shorter disk scale length than was adopted by these authors. 
Each of these leads to a change by a factor $\sim 1.4$ in the optical depth, 
but in opposite directions. We caution that our calculation of the optical 
depth towards Baade's Window is mainly for illustrative purpose since 
extending the disk profile to the Galactic center involves a substantial 
extrapolation from the data. By contrast, we expect that our optical 
depth estimate towards the LMC is reasonably accurate.

\section{CONCLUSIONS}

In this paper, we study M dwarfs from {\it Hubble Space Telescope} star
counts. The sample is about four times larger than that studied in 
Paper II.  This large sample considerably reduces the statistical errors 
which leads to the most precise determination of the M dwarf disk 
profiles to date. We also investigate the effects of systematic errors 
by using a modified color-magnitude relation which depends on Galactic 
height. The basic results are consistent with those in Paper I and II: 
the LF peaks at about $M_V=12$ and has a sharp drop toward $M_V=14$. The 
MF $d N / d \lg M \propto M^{\alpha}$ has a power-law index 
$\alpha=0.47\pm0.07$ for the solar-neighborhood CMR (1) and 
$\alpha =-0.10$ for the $z$-dependent CMR (2), both before the correction 
for binaries.  Our analysis favors a short scale length, 
$H = 2.75 \pm 0.16 {\rm (statistical)} \pm 0.25 {\rm (systematic)}$ kpc,
for the M dwarf disk population lying $\sim 2$ kpc above the plane.

\acknowledgments{Work by Z.Z., A.G., and S.S. was supported in part by 
grant AST~97-27520 from the NSF.  A.G. received additional support from 
Le Minist\`ere de L'\'Education Nationale de la Recherche et de la 
Technologie.}

\appendix{\bf APPENDIX: GROUND-BASED CALIBRATION OF HST WFPC2
PHOTOMETRY OF RED STARS}

For many astrophysical purposes, the use of standard 
Johnson-Cousins photometric system is necessary, as most of the 
ground-based results are expressed in this system. 
The {\it Hubble Space Telescope (HST)} Wide Field Planetary Camera 2 
(WFPC2) has its own set of filters with transmission profiles different 
from those of Johnson-Cousins system. A calibration between the two 
systems is required - in this case between WFPC2 F606W and F814W filters 
and standard  {\it V} and {\it I} filters, to which they roughly 
correspond. And in this particular case, the calibration has to be valid 
for the very red stars.

The most widely used such calibration is that by Holtzman et al.\ (1995, 
hereafter H95). They give a synthetic calibration for F606W based on 
convolutions of WFPC2 response curves with stellar spectra from the 
Bruzual, Persson, Gunn \& Stryker atlas (available from STScI ftp site.) 
While H95 do give an empirical calibration for F814W, it is derived using 
the ground-based observations of stars that were ``not very red''. 
The authors suggest that for red stars one should switch from the 
empirical transformations to the synthetic ones.

Bahcall et al.\ (1994, hereafter BFGK) give another synthetic calibration. 
They transformed F606W and F814W magnitudes into {\it V} and {\it I} 
using theoretical throughput and response curves and convolving them 
with Gunn \& Stryker (1983) standard spectra. A comparison between H95 
and BFGK transformations indicates some differences both in zero points 
and in color terms. There is an obvious need for an empirical ground-based 
calibration, which one would want in any case. Any such calibration should 
include red stars ($V-I>3$).

It should be noted that transformations as given by H95 apply to WFPC2 
CCDs after the 1994 Apr 23 cooldown from $-76^{\circ}$ C to $-88^{\circ}$ C, 
while BFGK transformations apply to observations made before the cooldown. 
The cooldown reduced the CTE (Charge Transfer Efficiency) effect and 
increased the QE (Quantum Efficiency).

We have performed a ground-based calibration based on stars observed in 
a portion of the Groth Strip, 28 contiguous WFPC2 fields. Our single CCD 
field spans approximately 9 original WFPC2 fields ($\sim 36$  
arcmin$^2$). We chose the Groth Strip because, in contrast to most WFPC2 
fields, the size of the field is well-matched to ground-based CCDs and 
because photometry of its stellar contents was already available 
(Paper II). We selected the field so as to include as many red ($V-I>3$) 
stars as possible.

The observations were performed on the Hiltner 2.4-m telescope at the 
MDM Observatory, using the Echelle CCD camera with SITE $2048\times 2048$ 
detector. We obtained a total of 7200 s in $V$ and 6000 s in $I$, in 
photometric conditions and $1''$ seeing. Photometry was performed to 
limiting magnitudes of $V\sim 25.1$ and $I\sim 23.6$. Photometry was 
reduced using observations of Landolt (1992) standards (as red as 
$V-I$=4.00 for $V$ and $V-I$=3.48 for $I$), 
and standard procedures in {\it IRAF}. Color terms needed to 
transform instrumental into standard magnitudes are: $-0.004\pm 0.005$ for 
$V$, and $0.031\pm 0.007$ for $I$.

WFPC2 images of the Groth Strip for which we have photometry were taken 
before the 1994 Apr 23 cooldown, during the time when the effects of CTE 
were more pronounced. The only reason that we use pre-cooldown 
observations, although the vast majority of observations were made after 
the cooldown, is that these observations are unique in terms of sky 
coverage. Deep fields observed after the cooldown are mostly single 
fields, therefore containing few of the red stars that are needed for 
the calibration. In order to correct for CTE, we used the CTE model with 
coefficients for $-76^{\circ}$ C, given by Stetson (1998). Our WFPC2 
images have a background of 50 $e^-$/pixel in F606W (exposure time 700 s) 
and 44 $e^-$/pixel in F814W (exposure time 1100 s). We used these 
background levels to evaluate CTE corrections for each individual star. 
Corrections turned out to be between 0.01 and 0.07 mag. According to 
Stetson (1998), these CTE corrections include position independent charge 
loss and should eliminate the apparent problem of ``long vs. short'' 
exposures.

Out of approximately 75 stars in our field that were measured on the 
WFPC2 images (8 of them red), we were able to detect some 50. This number 
was reduced to 34 stars in $V$ (4 of them red) and 40 stars in $I$ 
(all 8 red stars detected) after excluding stars that were contaminated 
by close-lying galaxies in the ground-based photometry (as cross-checked 
with {\it HST} images) and those that were saturated in the {\it HST} 
photometry.

BFGK introduce chip-to-chip zero-point offsets that are to be applied to 
each of the chips 2, 3 and 4 of WFPC2 (designated as $\delta_n$, $n=2,3,4$), 
because of different response curves of chips. We checked the validity 
of these offsets by measuring the mean differences between the BFGK 
{\it HST} $V$ and $I$ magnitudes and our ground-based magnitudes 
separately for each chip. Our results show that these offsets are not 
needed.  The likely reason for this is that the chip sensitivity 
differences are accounted for by the flat-fielding that is normalized to 
the WF3 chip {\it only}. Any remaining chip-to-chip offsets must be 
very small.

Now we present our best fits to the residuals between our ground-based 
data and the magnitudes from the H95 and BFGK transformations. To obtain 
the fits, we weighted the data points by the photometric errors scaled up 
to produce reduced $\chi^2\sim 1$. The zero points and errors were 
obtained from fits centered at $V-I=2$, in order to reduce correlation 
between the zero point and slope (color-term) errors.

Since our WFPC2 observations were made prior to the cooling, the zero 
points will not be the same if one uses observations at the lower 
temperature. In that case one should add $\Delta V_{CD} = 0.044$ and 
$\Delta I_{CD} = 0.007$ (Whitmore 1995). Otherwise, for pre-cooling 
observations, $\Delta V_{CD} =\Delta I_{CD}=0$.

The corrections to the H95 calibration are,
\begin{equation}
V = V_H + \Delta V_H,\qquad  
\Delta V_H=[-0.058(\pm 0.010)+\Delta V_{CD}]-0.037(\pm 0.013)[(V-I)-2],
\end{equation}
\begin{equation}
I = I_H + \Delta I_H,\qquad    
\Delta I_H=[0.004(\pm 0.012)+\Delta I_{CD}]+0.004(\pm 0.012)[(V-I)-2],
\end{equation}
where $V_H$ and $I_H$ are the transformations given by equation (9) in 
H95, with zero points and first and second order color-terms given in 
Table 10 of H95. 

Since H95 was derived with post-cooling observations, 
$\Delta V_{CD} = 0.044$ and $\Delta I_{CD} = 0.007$. The correction 
in $I$ is not significant. At our adopted midpoint, $V-I=2$, our 
{\it V}-band calibration is in good agreement with H95, 
$\Delta V_H = -0.014 \pm 0.010$. However, we find a difference in slope 
of $-0.037$, which means that the redder stars are slightly bluer.

The corrections to the BFGK calibration are,
\begin{equation}
V = V_B + \Delta V_B,\qquad 
\Delta V_B=[0.059(\pm 0.011)+\Delta V_{CD}]-0.040(\pm 0.014)[(V-I)-2],
\end{equation}
\begin{equation}
I = I_B + \Delta I_B,\qquad 
\Delta I_B=[0.007(\pm 0.014)+\Delta I_{CD}]+0.022(\pm 0.013)[(V-I)-2],
\end{equation}
where $V_B$ and $I_B$ are the transformations given by equation (2.1) 
in BFGK, but with $\delta_n=0$, as previously explained. The BFGK zero 
points and color-terms are given in the text of BFGK, below their 
equation (2.1).

Again, the corrections to $I_B$ are small. For $V_B$, we find the same 
slope offset, but the zero-point offset is much larger than for $V_H$.


\end{document}

%% file: tab1.tex
\renewcommand{\arraystretch}{.9} 
\setcounter{table}{0}
\begin{deluxetable}{ccrrccc}
\tablecaption{ Characteristics of the 95 New WFC2 Fields
              \label{tab:data1}}
\tablehead{ R.A. & Dec. & $l$ ($^\circ$)& $b$ ($^\circ$) & $I_{\rm max}$ & $I_{\rm min}$ & $\Omega$ \\ 
           (2000)&(2000)&               &                &               &  
         & (WFC2)}
\startdata

11 41 53.35 & $-$80 31 55.6 & 299.97 & $-$18.06 & 23.29 & 18.64 & 1.00 \\ 
21 51 08.26 & $+$29 00 00.3 &  81.96 & $-$19.18 & 23.98 & 19.45 & 1.00 \\ 
21 50 34.47 & $+$28 50 37.9 &  81.75 & $-$19.22 & 24.43 & 19.76 & 1.00 \\ 
21 51 21.87 & $+$28 44 05.9 &  81.81 & $-$19.42 &  23.90 & 19.76 & 1.00 \\ 
07 50 47.13 & $+$14 40 44.2 & 206.07 &  19.63 & 23.42 & 19.45 & 1.00 \\ 
04 32 07.19 & $+$17 57 17.1 & 179.16 & $-$20.06 & 21.60 &  16.90 & 1.00 \\ 
17 55 26.65 & $+$18 18 17.8 &  43.68 &  20.34 & 24.21 & 19.45 & 1.00 \\ 
04 21 37.47 & $+$19 31 48.8 & 176.17 & $-$20.95 & 23.28 & 18.84 & 1.00 \\ 
04 25 17.16 & $+$17 42 57.4 & 178.26 & $-$21.47 & 22.81 & 17.65 & 1.00 \\ 
06 15 45.38 & $+$70 57 40.8 & 143.38 &  22.71 & 22.42 & 17.65 & 1.00 \\ 
16 11 33.45 & $-$18 38 27.6 & 355.19 &  23.33 & 21.66 &  14.40 & 1.00 \\ 
04 09 41.07 & $+$17 04 19.5 & 176.15 & $-$24.71 & 23.47 & 18.84 & 1.00 \\ 
09 09 57.87 & $-$09 27 42.9 & 239.28 &  25.08 & 24.00 & 19.76 & 1.00 \\ 
13 22 12.29 & $-$36 41 22.6 & 309.78 &  25.77 & 23.53 & 17.89 & 1.00 \\ 
06 11 18.08 & $-$48 47 56.4 & 256.51 & $-$26.44 & 24.51 & 19.87 & 1.00 \\ 
19 38 10.83 & $-$46 19 50.2 & 352.31 & $-$27.02 & 24.09 & 19.84 & 1.00 \\ 
17 36 39.19 & $+$28 04 11.2 &  52.06 &  27.79 & 23.30 & 16.51 & 1.00 \\ 
17 36 23.05 & $+$28 01 01.1 &  51.98 &  27.83 & 23.29 & 18.64 & 1.00 \\ 
07 39 14.98 & $+$70 22 58.3 & 145.09 &  29.43 & 23.36 & 18.75 & 1.00 \\ 
19 40 41.24 & $-$69 15 56.3 & 326.37 & $-$29.58 & 24.16 & 19.45 & 1.00 \\ 
19 41 03.27 & $-$69 11 52.4 & 326.45 & $-$29.61 & 22.96 & 18.09 & 1.00 \\ 
13 39 41.64 & $-$31 34 14.5 & 314.82 &  30.18 & 23.41 & 18.64 & 1.00 \\ 
16 30 36.09 & $+$82 29 35.1 & 115.76 &  31.27 & 23.89 & 18.53 & 1.00 \\ 
08 11 58.81 & $+$75 00 30.6 & 139.45 &  31.31 & 23.52 & 18.64 & 1.00 \\ 
03 55 31.55 & $+$09 43 33.5 & 179.83 & $-$32.15 & 23.52 & 18.26 & 1.00 \\ 
11 21 28.23 & $-$24 55 15.8 & 278.31 &  33.61 & 23.28 & 18.64 & 1.00 \\ 
12 53 01.47 & $-$29 14 15.5 & 303.35 &  33.63 & 24.32 & 19.45 & 1.00 \\ 
17 12 23.18 & $+$33 35 41.6 &  56.72 &  34.25 & 24.25 & 19.45 & 1.00 \\ 
08 30 43.55 & $+$65 50 27.9 & 149.78 &  34.68 & 23.88 & 19.15 & 1.00 \\ 
04 55 54.74 & $-$21 55 09.5 & 221.88 & $-$34.68 & 22.81 & 18.09 & 1.00 \\ 
08 31 03.47 & $+$65 50 06.7 & 149.78 &  34.71 & 24.01 &  19.50 & 1.00 \\ 
03 05 16.46 & $+$17 28 18.5 & 162.78 & $-$34.84 & 23.35 & 18.75 & 1.00 \\ 
03 05 30.00 & $+$17 09 56.5 & 163.06 & $-$35.06 & 23.32 & 18.75 & 1.00 \\ 
08 54 16.58 & $+$20 03 37.6 & 206.82 &  35.69 & 23.83 & 19.45 & 1.00 \\ 
20 44 45.80 & $-$31 19 19.4 &  12.65 & $-$36.72 & 22.40 & 17.65 & 1.00 \\ 
10 05 46.00 & $-$07 41 24.5 & 247.87 &  36.90 & 24.28 & 19.45 & 1.00 \\ 
16 57 51.54 & $+$35 25 42.3 &  58.26 &  37.53 & 24.04 & 18.64 & 1.00 \\ 
14 41 53.03 & $-$17 18 38.0 & 337.15 &  38.11 & 24.13 & 19.73 & 1.00 \\ 
02 38 19.44 & $+$16 39 13.7 & 156.66 & $-$39.12 & 23.06 & 18.64 & 1.00 \\ 
16 01 12.86 & $+$05 36 02.5 &  16.22 &  40.06 & 24.09 & 18.64 & 1.00 \\ 
21 57 11.22 & $-$69 49 29.1 & 321.17 & $-$40.57 & 24.27 & 18.75 & 1.00 \\ 
16 09 12.21 & $+$65 32 00.0 &  98.33 &  40.90 & 24.06 & 19.80 & 1.00 \\ 
16 42 18.38 & $+$39 46 14.8 &  63.39 &  41.08 & 24.51 & 19.55 & 1.00 \\ 
04 07 31.31 & $-$12 08 33.2 & 204.82 & $-$41.80 & 22.53 & 17.65 & 1.00 \\ 
14 49 56.96 & $-$10 06 03.5 & 344.63 &  42.97 & 23.92 & 19.76 & 1.00 \\ 
16 24 12.97 & $+$48 09 08.8 &  74.92 &  44.06 & 23.46 & 18.75 & 1.00 \\ 
01 15 51.87 & $+$16 41 57.3 & 131.33 & $-$45.77 & 22.39 & 17.89 & 1.00 \\ 
00 17 11.03 & $+$15 48 54.5 & 110.97 & $-$46.26 & 23.13 & 18.93 & 1.00 \\
\enddata
\end{deluxetable}

\setcounter{table}{0}
\begin{deluxetable}{ccrrccc}
\tablecaption{ Characteristics of the 95 New WFC2 Fields --- Continued
              \label{tab:data2}}
\tablehead{ R.A. & Dec. & $l$ ($^\circ$)& $b$ ($^\circ$) & $I_{\rm max}$ & $I_{\rm min}$ & $\Omega$ \\           
           (2000)&(2000)&               &                &               &  
         & (WFC2)}
\startdata

09 39 33.74 & $+$41 32 46.1 & 179.88 &  48.45 & 23.91 & 19.22 & 1.00 \\ 
10 56 59.04 & $-$03 35 27.8 & 256.54 &  48.69 & 23.25 & 18.64 & 1.00 \\ 
15 43 24.56 & $+$53 52 45.9 &  85.35 &  48.77 &  24.30 & 19.84 & 1.00 \\ 
22 32 55.80 & $-$60 33 01.1 & 328.25 & $-$49.21 & 23.97 &  19.40 & 1.00 \\ 
00 53 36.12 & $+$12 49 46.1 & 123.75 & $-$50.04 & 23.87 & 19.45 & 1.00 \\ 
00 53 23.16 & $+$12 33 57.7 & 123.68 & $-$50.30 & 23.54 & 19.45 & 1.00 \\ 
14 00 13.84 & $+$62 33 42.4 & 109.91 &  52.79 & 23.71 & 18.84 & 1.00 \\ 
13 59 59.37 & $+$62 33 42.5 & 109.95 &  52.81 & 23.65 & 18.84 & 0.95 \\ 
14 00 06.58 & $+$62 31 06.5 & 109.90 &  52.84 & 23.67 & 18.84 & 0.95 \\ 
13 59 52.14 & $+$62 31 06.6 & 109.94 &  52.85 & 23.71 & 18.84 & 0.92 \\ 
13 59 37.69 & $+$62 31 06.5 & 109.98 &  52.86 & 23.71 & 18.84 & 0.93 \\ 
14 00 13.77 & $+$62 28 30.4 & 109.85 &  52.87 & 23.67 & 18.84 & 1.00 \\ 
13 59 23.24 & $+$62 31 06.4 & 110.03 &  52.87 & 23.67 & 18.84 & 0.94 \\ 
13 59 59.35 & $+$62 28 30.5 & 109.89 &  52.88 & 23.63 & 18.84 & 0.96 \\ 
13 59 44.92 & $+$62 28 30.5 & 109.93 &  52.90 & 23.62 & 18.84 & 0.96 \\ 
13 59 30.50 & $+$62 28 30.4 & 109.97 &  52.91 &  23.70 & 18.84 & 0.96 \\ 
10 24 51.70 & $+$47 05 33.1 & 168.18 &  55.10 & 23.18 & 18.64 & 1.00 \\ 
15 19 54.79 & $+$23 44 53.3 &  35.59 &  56.43 & 23.92 & 19.28 & 1.00 \\ 
14 41 00.08 & $+$53 26 59.2 &  92.90 &  56.81 & 23.34 & 18.64 & 1.00 \\ 
01 24 41.93 & $+$03 51 26.1 & 138.73 & $-$57.99 & 24.19 & 19.72 & 1.00 \\ 
13 36 17.10 & $-$00 52 02.8 & 325.81 &  60.00 & 23.87 & 17.34 & 1.00 \\ 
12 36 39.59 & $-$00 41 58.0 & 295.06 &  61.95 & 23.64 & 18.26 & 1.00 \\ 
14 42 30.88 & $+$35 24 22.1 &  59.12 &  64.96 & 22.84 & 18.09 & 1.00 \\ 
00 50 31.47 & $-$52 09 55.0 & 303.26 & $-$64.96 &  22.80 & 17.18 & 1.00 \\ 
00 50 32.91 & $-$52 07 18.5 & 303.26 & $-$65.01 & 23.33 & 18.64 & 0.97 \\ 
00 50 12.11 & $-$52 03 51.7 & 303.38 & $-$65.06 & 23.69 & 19.08 & 1.00 \\ 
12 34 08.55 & $+$02 44 33.8 & 292.56 &  65.27 & 23.63 & 18.75 & 1.00 \\ 
13 53 29.96 & $+$48 32 55.8 &  97.73 &  65.43 & 24.14 & 19.76 & 1.00 \\ 
11 46 02.35 & $+$47 34 03.4 & 150.64 &  65.89 & 22.54 & 17.65 & 1.00 \\ 
11 46 14.53 & $+$47 33 53.6 & 150.58 &  65.91 & 22.83 & 18.09 & 1.00 \\ 
12 17 54.57 & $+$50 12 11.8 & 136.22 &  66.05 & 24.24 & 19.87 & 1.00 \\ 
11 16 27.41 & $+$18 05 42.7 & 230.38 &  66.35 & 23.91 & 19.28 & 1.00 \\ 
14 35 33.34 & $+$25 18 09.0 &  34.37 &  66.62 & 24.06 & 18.64 & 1.00 \\ 
11 48 21.38 & $+$10 50 03.1 & 257.58 &  67.96 & 23.16 & 18.09 & 1.00 \\ 
14 04 28.90 & $+$43 19 12.3 &  85.29 &  68.08 & 24.34 & 19.80 & 1.00 \\ 
11 48 49.73 & $+$10 55 05.9 & 257.68 &  68.10 & 23.94 & 18.26 & 1.00 \\ 
11 48 50.97 & $+$10 57 56.2 & 257.61 &  68.14 & 24.13 & 19.55 & 1.00 \\ 
12 27 45.96 & $+$44 07 58.1 & 137.05 &  72.34 & 22.96 & 18.26 & 1.00 \\ 
12 10 33.65 & $+$39 28 58.7 & 154.88 &  75.00 & 23.93 & 18.26 & 1.00 \\ 
00 15 47.24 & $-$16 19 06.3 &  83.73 & $-$76.39 & 23.27 & 18.64 & 1.00 \\ 
00 15 55.35 & $-$16 18 06.2 &  83.88 & $-$76.40 & 23.12 & 18.64 & 1.00 \\ 
11 50 29.28 & $+$28 48 33.4 & 202.26 &  76.45 & 24.11 & 19.45 & 1.00 \\ 
13 16 28.86 & $+$36 27 16.1 &  94.83 &  79.27 & 22.81 & 18.09 & 1.00 \\ 
13 24 49.72 & $+$30 58 36.0 &  62.72 &  81.75 & 24.87 & 19.55 & 1.00 \\ 
12 47 47.30 & $+$34 32 57.3 & 128.73 &  82.54 & 23.25 & 18.64 & 1.00 \\ 
00 24 54.07 & $-$27 16 17.7 &  30.02 & $-$84.10 & 24.14 & 19.76 & 1.00 \\ 
13 00 23.85 & $+$28 20 06.1 &  64.77 &  87.68 & 23.31 & 18.64 & 1.00 \\ 
\enddata
\end{deluxetable}